\documentclass[aps,prl,twocolumn,superscriptaddress,10pt]{revtex4-1}
\pdfoutput=1
\usepackage{amsmath,bm,amstext,amssymb,mathtools}

\usepackage{hyperref}
\usepackage{cleveref}
\usepackage{graphicx}

\renewcommand{\k}{{\bm{k}}}
\newcommand{\B}{\textnormal{B}}
\newcommand{\p}{\bm{p}}
\newcommand{\q}{\bm{q}}
\newcommand{\0}{\bm{0}}
\newcommand{\ab}{a_{\textnormal{B}}}
\newcommand{\mb}{m_{\textnormal{B}}}

\newcommand{\ek}{\epsilon_{\k}}

\newcommand{\T}{\mathcal{T}}
\newcommand{\V}{\mathcal{V}}
\newcommand{\dif}{\textnormal{d}}

\newcommand{\nn}{\nonumber}

\begin{document}
\title{Quasiparticle Properties of a Mobile Impurity in a Bose-Einstein Condensate}
\author{Rasmus S\o gaard Christensen}
\affiliation{Department of Physics and Astronomy,  Aarhus University, DK-8000 Aarhus C, Denmark}
\author{Jesper Levinsen}
\affiliation{School of Physics and Astronomy, Monash University, Victoria 3800,
  Australia }
\affiliation{Aarhus Institute of Advanced Studies, Aarhus University,
  DK-8000 Aarhus C, Denmark}
\author{Georg M. Bruun}
\affiliation{Department of Physics and Astronomy, Aarhus University, DK-8000 Aarhus C, Denmark}
\date{\today}

\begin{abstract}
  We develop a systematic perturbation theory for the quasiparticle
  properties of a single impurity immersed in a Bose-Einstein
  condensate. Analytical results are derived for the impurity
  energy, the effective mass, and residue to third order in the
  impurity-boson scattering length. The energy is shown to depend
  logarithmically on the scattering length to third order, whereas the
  residue and effective mass are given by analytical power series.   
  When the boson-boson scattering length equals the boson-impurity scattering length,  the energy   
   has the same structure  as that of a weakly interacting Bose gas, including terms of the Lee-Huang-Yang and fourth order logarithmic form.  
    Our results, which cannot be  obtained within the canonical Fr{\"o}hlich model of an impurity
  interacting with 
  phonons, provide valuable benchmarks for
   many-body theories and for experiments.
\end{abstract}
\pacs{}
\maketitle

The problem of an impurity interacting with a reservoir with a
continuous set of degrees of freedom plays a fundamental role 
in our understanding of many-body quantum systems. Landau and Pekar
famously demonstrated that electrons in a dielectric medium become
dressed by phonons forming a quasiparticle termed a
polaron~\cite{Landau1933,Pekar1946}.  Other examples of mobile
impurities include helium-3 mixed with
helium-4~\cite{BaymPethick1991book} and $\Lambda$ particles in nuclear
matter~\cite{Bishop1973}.  Static impurities give rise to the Anderson
orthogonality catastrophe~\cite{Anderson1967} and the Kondo
effect~\cite{Kondo1964}. With the creation of two-component atomic
gases characterized by an unrivaled experimental flexibility, the
impurity problem can now be studied more systematically and from a
broader perspective. While focus has mostly been on impurities in a
Fermi sea (the Fermi
polaron)~\cite{Schirotzek2009,Kohstall2012,Koschorreck2012}, there
have been some experiments on impurity atoms in a Bose
gas~\cite{Catani2012,Fukuhara2013,Palzer2009,Chikkatur2000,Spethmann2012,Scelle2013}. With
the recent identification of Feshbach resonances in
Bose-Fermi~\cite{Wu2012, Heo2012,Cumby2013} and
Bose-Bose~\cite{Roati2007,Pilch2009} mixtures, the study of impurity
physics in a Bose-Einstein condensate (BEC) with a tunable interaction is now within reach.

The impurity problem provides an ideal setting for testing many-body
theories, and it has yielded fundamental insights for the Fermi
polaron~\cite{Massignan_Zaccanti_Bruun}. In the case of an impurity
atom in a BEC---the Bose polaron---most studies have either used
mean-field theory to study
self-localization~\cite{Astrakharchik2004,Cucchietti2006,Kalas2006,Bruderer2008}
and time dependence for weak coupling~\cite{Volosniev2015}, or
employed an effective Fr\"ohlich model which is solved using various
many-body
techniques~\cite{Huang2009,Tempere2009,Casteels2014,Shashi2014,Grusdt2014,Vlietinck2014}. The
Fr\"ohlich model, however, ignores interaction terms important even
for weak coupling, as we shall demonstrate. The correct microscopic
Hamiltonian has been used in a field theoretic approach, selectively
summing ladder diagrams~\cite{Rath2013}, and in a variational
approach~\cite{Li2014}.

Since the impurity problem is so useful as a theoretical testing
ground, it is important to have a quantitatively reliable theory,
which can serve as a benchmark for other many-body theories and for experiments in the
weak coupling regime.  Here, we provide such an accurate theory 
 by developing a systematic perturbation expansion for
the impurity self-energy to \emph{third order} in the impurity-boson
scattering length $a$. The small parameter of this expansion is $a/\xi$, with  $\xi$ being
the BEC coherence length. 
Also,  $a^2/\xi a_\B$ has to be small for the polaron to be well defined, where  $a_{\B}$ is the boson-boson scattering
length. We derive analytical results for the zero temperature quasiparticle
properties of the polaron, showing that 
the energy contains a logarithmic term $\ln(a^*/\xi)a^3/\xi^3$, where
$a^*\sim\max(a,a_\B)$. When $a=\ab$, the perturbative expression for the energy has  the same form as the 
celebrated result for a weakly interacting Bose gas~\cite{LeeYang1957,LeeHuangYang1957,Wu1959,Hugenholtz1959,Sawada1959}.  The quasiparticle residue and the effective mass are, on the other hand, given by analytic power series up to  $a^3/\xi^3$.
We use a ${\mathcal T}$-matrix approach, a technique which has previously been
employed for the Fermi polaron to sum diagrams to  high order 
with Monte Carlo methods~\cite{Prokofev2008a,Prokofev2008b,Vlietinck2013}.
Our approach uses the correct microscopic Hamiltonian
rather than the Fr\"ohlich model used in Refs.~\cite{Huang2009,Tempere2009,Casteels2014,Shashi2014,Grusdt2014,Vlietinck2014}, 
as the latter is symmetric with respect to the sign of
the interaction and therefore  only contains 
even powers in perturbation theory  beyond the mean-field shift.  Also, the field theory~\cite{Rath2013} and the variational approach~\cite{Li2014} 
 miss terms at second order.

We consider an impurity of mass $m$ immersed in a BEC of particles
with mass $m_{\B}$. The Hamiltonian is
\begin{align}
   H=&\sum_\k\ek^{\B \vphantom{\dagger}}a^\dagger_\k a_\k^{\vphantom{\dagger}}
   			+ \frac{1}{2\V} \sum_{\k,\k',\q}
V_{\B}(q) a^\dagger_{{\k}+{\q}}a_{{\k}'-\q}^\dagger
 a_{{\k}'}^{\vphantom{\dagger}} a_{{\k}}^{\vphantom{\dagger}}
\nonumber\\
& +\sum_{{\k}}\epsilon_{\k}^{\vphantom{\dagger}} c^\dagger_{{\k}}c_{{\k}}^{\vphantom{\dagger}}
 +\frac{1}{\V} \sum_{{\k},{\k}',{\q}}V(q)c_{{\k}+\q}^\dagger a^\dagger_{{\k}'-{\q}} a_{{\k}'}^{\vphantom{\dagger}} c_{{\k}}^{\vphantom{\dagger}},
\end{align}
where $a_{{\k}}$ and $c_{{\k}}$ removes a boson and an
impurity, respectively, with momentum ${\k}$,
$\ek^{\B}=k^2/2m_{\B}$ and $\ek={k^2}/{2m}$ are the free
dispersions, and $\V$ is the system volume. The boson-boson
$V_{\B}(q)$ and boson-impurity interaction $V(q)$ are  assumed
to be short ranged, and they give the usual zero energy scattering matrices ${\T}_{\B}=4\pi a_{\B}/m_{\B}$ and
${\T}_v=2\pi a/m_{r}$ respectively, with $m_r=m_{\B}m/(m_{\B}+m)$ the
reduced mass,  see the Supplemental Material
\cite{supmat}.  We work in units where $\hbar=k_B=1$.

\begin{figure}[tbp]
\centering
\includegraphics[width=\columnwidth]{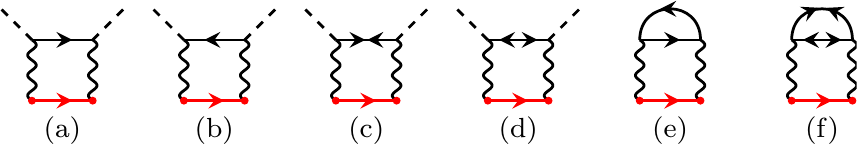}
\caption{
Second order diagrams for the self-energy. The upper solid
  black lines are the normal, $G_{11}$, and anomalous, $G_{12}$ and
  $G_{21}$, boson propagators. The dashed lines are particles emitted from
  or absorbed into the BEC, the bottom red lines are the impurity
  propagators, and the external impurity propagators are attached to
  the red dots. The wavy lines denote the impurity-boson vacuum
  scattering matrix $\T_v$.
 All diagrams come from ladder-type diagrams similar to those arising in
 	Eq.~\ref{Tmatrix}: The pair propagator $\Pi_{11}$ appears in (a) and (e),
 	$\Pi_{22}$ in (b),
 	$\Pi_{12}$ in (c), 
 	and $\Pi_{21}$ in (d) and (f).
}
\label{2orderdiagrams}
\end{figure}

\begin{figure}[tbp]
\centering
	\includegraphics[width=\columnwidth]{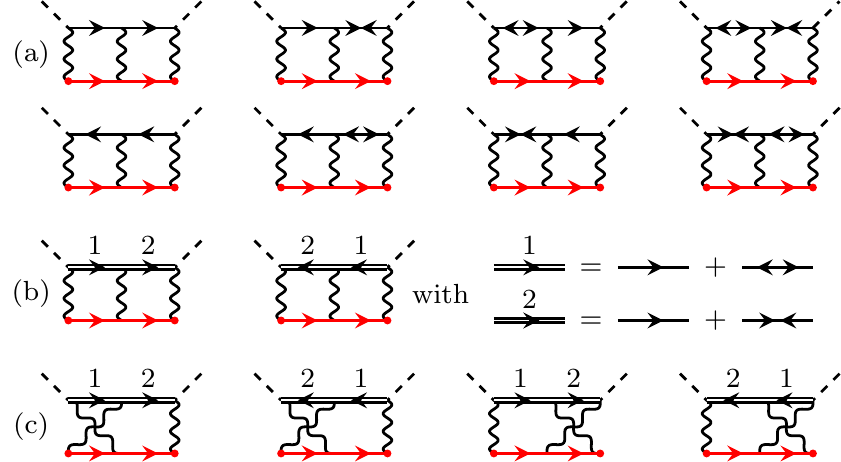}
	\caption{
	Third order diagrams for the self energy:
	(a) $\Sigma_{3b}$, 
	(b) $\Sigma_{3b}$ expressed using the  propagator $G(p)$, and 
	(c) $\Sigma_{3c}$.
	}
	\label{3orderdiagrams}
\end{figure}

The BEC is assumed to be weakly interacting such that it can be
described by Bogoliubov theory, {\em i.e.} $n_0\ab^3\ll1$ with $n_0$
being the condensate density. We define the imaginary time Bose Green's
functions as
$G_{11}({\k},\tau)=-\langle T_\tau\{a_{\k}^{\vphantom{\dagger}}(\tau),a_{\k}^\dagger(0)\}\rangle$,
$G_{12}({\k},\tau)=-\langle T_\tau\{a_{-\k}(\tau),a_{\k}(0)\}\rangle$
and
$G_{21}({\k},\tau)=-\langle T_\tau\{a^\dagger_{\k}(\tau), a^\dagger_{-\k}(0)\}\rangle$,
where $T_\tau$ denotes time ordering.  The Fourier transforms are
\begin{align}
G_{11}(\k,z)=\frac{u^2_\k}{z-E_\k}-\frac{v^2_\k}{z+E_\k}, \hspace{0.5cm}
G_{12}(\k,z)=\frac{u_\k v_\k}{E_\k^2-z^2},
\nn
\end{align}
with $u_\k^2=(\xi_\k/E_\k+1)/2$, $v_\k^2=(\xi_\k/E_\k-1)/2$,
$E_\k=\sqrt{\xi_\k^2-{\T}_{\B}^2n_0^2}$,
$\xi_\k=\epsilon_{\k}^{\B}+{\T}_{\B}n_0$, and
$G_{21}(\k,z)=G_{12}(\k,z)$.
Here,$z=i2sT$ is a Bose Matsubara frequency,
$s$ is an integer, and $T$ is the temperature,

\emph{Perturbation series.---}%
Our aim is to develop a systematic perturbation theory in powers of the impurity-boson scattering length
for the quasiparticle properties of the impurity. To this end, we write down all diagrams for the impurity self-energy up to third order in 
the bare interaction $V(q)$. We then formally replace $V(q)$ with $\T_v$ in each diagram. Diagrams which contain the
simultaneous forward propagation of an impurity and a boson (the pair propagator), such as the second order diagram in 
Fig.~\ref{2orderdiagrams}(a) and the three first third order diagrams
in Fig.~\ref{3orderdiagrams}(a), can be thought of as coming from the expansion
\begin{align}
\T( p) =
\frac{ \T_v }{1 - \T_v \Pi_{11}( p)}=\T_v^{\vphantom{2}}+\T_v^2 \, \Pi_{11}(p)+\cdots
\label{Tmatrix}
\end{align}
of the ladder approximation for the impurity-boson scattering matrix
in the BEC~\cite{Fetter}.  Here, $\Pi_{11}( p)$ denotes the pair
propagator regularized by subtracting the vacuum scattering already
contained in $\T_v$. We use the shorthand notation $p =(\p,\omega)$
having analytically continued to real energy
$z\rightarrow\omega+i0_+$. The perturbative expansion (\ref{Tmatrix})
is  convergent only for small $\T_v\Pi_{11}(p)$, i.e.
small $a$, whereas the full frequency dependence of $\T(p)$ has to be
retained when $a$ is large. This approach, which is detailed in the
Supplemental Material, yields a perturbation series in $a$ for the
self-energy: $\Sigma(p)=\Sigma_1(p)+\Sigma_2(p)+\Sigma_3(p)+\cdots$,
where $\Sigma_n$ contains diagrams of order $a^n$. The first order
term is the mean-field energy shift $\T_vn$ with $n$ being the
total boson density. We now evaluate the next two terms.

\emph{Second order.---}%
We now evaluate the next two terms.
Second  order
.
—
The six second order self-energy diagrams are shown in Fig.~\ref{2orderdiagrams}.
Fig.~\ref{2orderdiagrams}(a) is given by $n_0\T_v^2\Pi_{11}$
and---together with Fig.~\ref{2orderdiagrams}(e)---comes from
expanding the $\T$-matrix given by
Eq.~(\ref{Tmatrix}) to second order
in $a$ inside the ladder approximation to the self-energy. The three
diagrams in Figs.~\ref{2orderdiagrams}(b)--(d) are given by
$n_0\T_v^2\Pi_{22}$, $n_0\T_v^2\Pi_{21}$, and
$n_0\T_v^2\Pi_{12}$, respectively, with $\Pi_{21}=\Pi_{12}$
  representing the anomalous propagators and $\Pi_{22}$ the particle-hole
propagator\
\cite{supmat}. Apart from ladder summations inside
$\T_v$, the first four diagrams in Fig.~\ref{2orderdiagrams}
only contain scattering of bosons into or out of the BEC, and they can
in fact be obtained from  the Fr\"ohlich
model if one replaces $g\to \T_v$ by hand~\cite{Huang2009}. Figures~\ref{2orderdiagrams}(e) and (f)
contain vertices where both the in- and outgoing bosons are outside
the BEC, and they are not included in the Fr\"ohlich model.  They are,
however suppressed by a factor $(n_0\ab^3)^{1/2}$, and since we only
consider terms to lowest order in $n_0\ab^3$, these two diagrams will
be ignored. Likewise, we do not distinguish between $n_0$ and $n$ in
the following.

In total, the second order self-energy is
\begin{align}
\Sigma_2(p)=n_0\T_v^2[\Pi_{11}(p)+2\Pi_{12}(p)+\Pi_{22}(p)].
\label{eqn:Sigma2}
\end{align}
For $T=0$ and $p=(\0,0)$, $\Pi_{11}(p)$, $\Pi_{12}(p)$, and $\Pi_{22}(p)$ may be found analytically \cite{supmat} to yield
 \begin{align}
   \Sigma_2(0)=A(\alpha)\frac{2\pi n_0}{m_r}\frac{a^2}{\xi},
\label{SelfSecond1}
\end{align}
where $\alpha\equiv m/\mb$ is the mass ratio and  $\xi=1/\sqrt{8\pi na_{\B}}$ is the coherence length of
the BEC.  We have $A(1)=8\sqrt{2}/3\pi$, and  
 $A(\alpha)$ for a general mass ratio is given analytically in the 
Supplemental Material~\cite{supmat}.

\emph{Third order.---}We now consider the diagrams for
$\Sigma_3(p)$. They can be divided into three different classes. The
first, denoted $\Sigma_{3a}(p)$, is obtained by inserting first order self-energies
$\T_vn_0$ for the impurity propagators in the second order
diagrams depicted in Figs.~\ref{2orderdiagrams}(a)--(d). This 
yields four diagrams which are easily evaluated. As we shall see,
however, 
these self-energy insertions are canceled by a similar first order
shift $\T_vn_0$ in the impurity energy, which must be
inserted in the second order diagrams.

The second class of third order diagrams consists of the eight
``ladder'' diagrams depicted in Fig.~\ref{3orderdiagrams}(a). They are
easily expressed in terms of the two-particle propagators
$\Pi_{ij}$. Using the effective propagator $G=G_1=G_2$, with
$G_1=G_{11}+G_{12}$ and $G_2=G_{11}+G_{21}$, they can be reduced to
the two diagrams shown in Fig.~\ref{3orderdiagrams}(b). Their sum is
\begin{align}
  \Sigma_{3b}=n_0{\mathcal
  T}_v^3[(\Pi_{11}+\Pi_{12})^2+(\Pi_{22}+\Pi_{12})^2],
\label{Ladder3}
\end{align}
where we have suppressed the momentum and frequency dependence for notational simplicity. 

The third class of diagrams are those where either the first or the
last two interaction lines are crossed. There are 16 such
diagrams, but using the propagator $G(p)$ they can be reduced to the
four terms depicted in Fig.~\ref{3orderdiagrams}(c), which constitutes
a major simplification.  Evaluating the Matsubara sums and
specializing to $T=0$ yields for the sum of the crossed diagrams
\begin{align}
  \Sigma_{3c}(p) =& \, 2 n_0 \T_v^3 \int \! \dif\check k \!
                    \left\{	\frac{ v_{\k}^{2} - u_{\k} v_{\k} }{ z - E_{\k} - \epsilon_{\bm{\k'}} }
                    \left[ \Pi_{11}(k') + \Pi_{12}(k') \right]\right. 
                    \nonumber\\ & %
                                  \left.	+	\frac{ u_{\k}^{2} - u_{\k} v_{\k} }{ z - E_{\k} - \epsilon_{\bm{\k'}} }
                                  \left[ \Pi_{22}(k') + \Pi_{12}(k') \right]
                                  \right\},
\label{Crossed}
\end{align}
with $k'\equiv (\bm{\p +\k}, z - E_{k})$.  

Equation \eqref{Crossed} is, in fact, ultraviolet divergent, the
offending terms being $u_\k v_\k\Pi_{11}(\bm{k'})$ and
$u_\k^2 \Pi_{12}(\bm{k'})$. As shown in the Supplemental Material
\cite{supmat}, these two terms give rise to a $1/k$ behavior of the
integrand for $\Sigma_{3c}$ for large $k$, which thus appears to be
logarithmically divergent.  However, this should not cause us too much
worry. First, the integrand is well behaved at low momenta, where a
natural lower cutoff is provided by $k\sim 1/\xi$, below which the boson dispersion becomes linear.   Second, the ultraviolet divergence is a
consequence of the fact that we have assumed a constant scattering
matrix ${\T}_{\B}$. 
Retaining the energy dependence of
${\T}_{\B}$ would result in an ultraviolet cutoff
$\sim 1/a_{\B}$.  Likewise, replacing ${\T}_v$ by the full
energy dependent scattering matrix ${\T}$ in the diagrams for
$\Sigma_{3c}$ gives a cutoff $\sim 1/a$. We can therefore write
\begin{align}
\Sigma_3(0)=B(\alpha)\frac{2\pi n_0}{m_r}\frac{a^3}{\xi^2}\ln(a^*/\xi)+{\mathcal O}(a^3/\xi^3),
\label{SelfThird1}
\end{align}
with $a^*\equiv\max(a,a_\B)$. For the equal mass case we have $B(1)=2/3-\sqrt{3}/\pi$, and the analytic expression 
for general $\alpha$ is given in the Supplemental Material~\cite{supmat}.
 Note that since $\Sigma_{3b}(p)\sim{\mathcal O}(a^3/\xi^3)$, it does not
contribute to the self-energy to the order stated in
Eq.~(\ref{SelfThird1}) for $a > a_{\B}$. It does, however, contribute to the
quasiparticle residue and effective mass, as we shall see below.

\emph{Quasiparticle energy.---}Having evaluated the 28 third
order diagrams, we can now present a perturbative expression for the
polaron energy given by the solution of
$E(\bm{p})=\bm{p}^2/2m+\Sigma\left[\p,E(\bm{p})\right]$.  From
Eqs.~(\ref{SelfSecond1}) and (\ref{SelfThird1}), we obtain for $\p=\0$
and $T=0$
\begin{align}
  \frac{E(0)}{\Omega} = \frac{a}{\xi}+A(\alpha)\frac{a^2}{\xi^2}+
  B(\alpha)\frac{a^3}{\xi^3}\ln(a^*/\xi)
\label{eq:mu}
\end{align}
where $\Omega=2\pi n_0 \xi/m_r$ is the mean-field polaron energy for
$a=\xi$.  Equation \eqref{eq:mu} gives the polaron energy to order
$\ln(a^*/\xi)a^3/\xi^3$ and is one of our main results. We see that
the small parameter of the perturbation series is $a/\xi$. The second
order term agrees with that obtained using the Fr\"ohlich
Hamiltonian~\cite{Casteels2014}.  As we can see from
Fig.~\ref{3orderdiagrams}(b) however, the third order logarithmic term
comes from scattering events where both bosons are excited out of the
BEC. These are precisely the processes ignored by the Fr\"ohlich
model, which therefore incorrectly predicts a vanishing third order
term.  On the other hand, at fourth order in $a$ the Fr\"ohlich model
has been shown to also have a logarithmic contribution~\cite{Grusdt2014}.


%
%



Interestingly, when $a=\ab$, Eq.\ \eqref{eq:mu} has the same structure as the famous result for the  energy  of a weakly interacting 
Bose gas: schematically $E\sim a[1+(na^3)^{3/2}+na^3\ln(na^3)]$~\cite{Fetter}. Since it is difficult to measure the bulk energy 
of a Bose gas, the Lee-Huang-Yang  $(na^3)^{3/2}$ term~\cite{LeeYang1957,LeeHuangYang1957} 
 was measured only recently~\cite{Navon2010}, whereas the logarithmic correction~\cite{Wu1959,Hugenholtz1959,Sawada1959} has never been detected. 
The  energy of an impurity atom has, however, been measured accurately using radio-frequency (rf) 
spectroscopy~\cite{Schirotzek2009,Kohstall2012,Koschorreck2012}, and  our  result therefore suggests a way to measure 
beyond mean-field effects in Bose gases including logarithmic corrections for the first time.

\emph{Quasiparticle residue and effective mass.---}The quasiparticle
residue is given by $Z^{-1}=1-\partial_\omega\Sigma$. For zero
momentum, we obtain to third order
\begin{align}
& \hspace{-2mm} 1-Z^{-1}
=\partial_\omega\!\!\left.\Sigma_2(\0,\omega)\right|_{{\mathcal
   T}_vn_0}+\partial_\omega\!\!\left.\Sigma_3(\0,\omega)\right|_{0}
\nonumber \\
& \hspace{4mm} =
\partial_\omega\!\!\left.\Sigma_2(\0,\omega)\right|_{0}+
  \partial_\omega\left.\!\![\Sigma_{3b}(\0,\omega)+\Sigma_{3c}(\0,\omega)]\right|_{0}.
 \label{Residue}
\end{align}
The second line follows from inserting the first order shift $\omega={\T}_vn_0$ into the second order 
 self-energy. When expanding in  ${\T}_vn_0$, this  yields a  third order term, which cancels the third order diagrams $\Sigma_{3a}$; see the Supplemental Material
\cite{supmat}. Contrary to the case
of the  energy, the self-energy term $\Sigma_{3b}$
contributes to the residue since $\partial_\omega\Sigma_{3b}$ and
$\partial_\omega\Sigma_{3c}$ are of the same order. We can evaluate $\partial_\omega\Sigma_2$ and $\partial_\omega\Sigma_{3b}$  analytically,
 whereas
$\partial_\omega\Sigma_{3c}$ has to be calculated numerically. We obtain
\begin{equation}
Z^{-1}=1 + C(\alpha)\frac{a^2}{a_\B\xi} + D(\alpha)\frac{a^3}{a_\B\xi^2},
\label{ResidueFinal}
\end{equation}
where $C(\alpha)$ and $D(\alpha)$ are given in the Supplemental
Material \cite{supmat}. For $m=m_\text{B}$, we have $C(1)=2\sqrt2/3\pi$
and $D(1)\approx0.64$. Equation (\ref{ResidueFinal}) explicitly shows that the polaron 
is  well defined only for $a^2/a_\text{B}\xi\ll 1$.

For an ideal BEC with $a_\B=0$, we have  $\xi\rightarrow\infty$, and it follows from  Eq.~\eqref{eq:mu} 
 that there are no corrections to the  mean-field energy up to third order in $a$. 
However, in this limit Eq.~\eqref{ResidueFinal} predicts $Z=0$ so that there 
 is no well-defined quasiparticle, signaling a breakdown of  
perturbation theory. 
 The reason is that the energy of the impurity atom is right at the threshold of the particle-hole continuum of the 
  BEC, giving rise to a square root energy dependence of the self-energy and thus zero  residue as explained in the Supplemental Material
\cite{supmat}. Equivalently, 
 Landau's critical velocity $c=(4\pi \ab n)^{1/2}/\mb$ above which the polaron decays through momentum relaxation,  
 is zero for a noninteracting BEC.

 The effective mass of the quasiparticle is obtained from
$m/m^*=Z(1+2m\partial_{p^2}\Sigma)$. Following steps analogous to the
calculation of $Z$, we obtain
\begin{align}
	\frac{m^{\mathrlap{*}}}{m}=1+F(\alpha)\frac{a^2}{a_\B\xi}+G(\alpha)\frac{a^3}{a_\B\xi^2},
	\label{eqn:mstar_final}
\end{align}
where $F(\alpha)$ and $G(\alpha)$ are given in the Supplemental
Material \cite{supmat}. For the equal mass case, we have
$F(1)=16\sqrt{2}/45\pi$ and $G(1)\approx 0.37$. Our result for
$F(\alpha)$ matches that of Ref.~\cite{Casteels2014}.

\begin{figure}[tbp]
	\centering
	\includegraphics[width=\columnwidth]{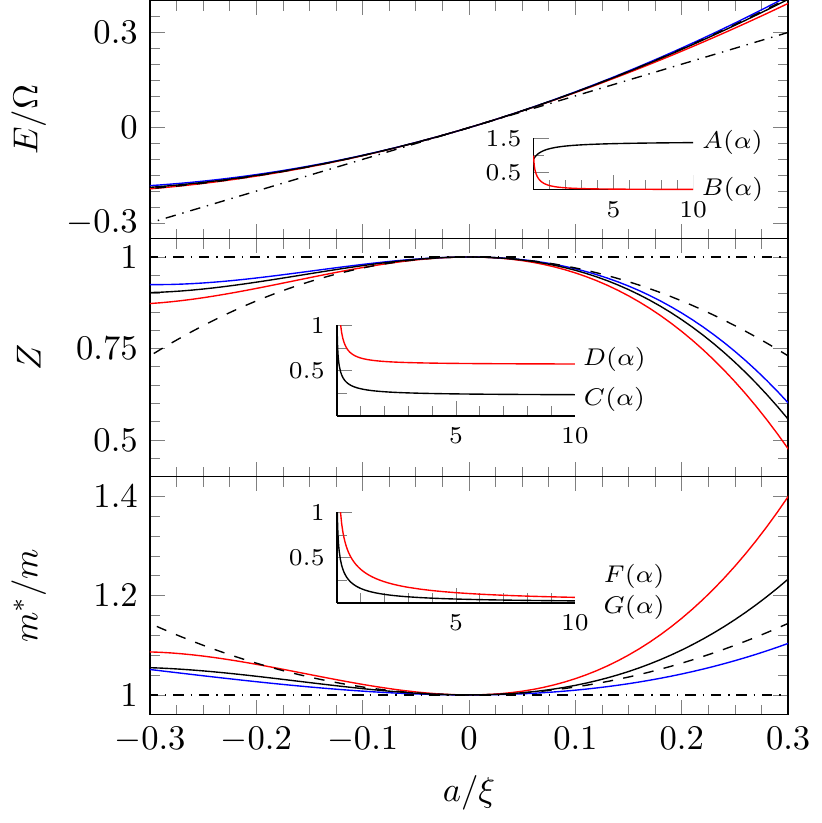}
	\caption{The polaron energy $E$, residue $Z$ and effective
          mass $m^*$ as given by Eqs.~\eqref{eq:mu},
          \eqref{ResidueFinal}, and \eqref{eqn:mstar_final} for
          $a_{\B}/\xi = 0.1$. We show the first order term
          (dash-dotted lines), second order (dashed lines) and third order
          (solid lines) for equal masses (black lines). For $\alpha=39/87$ (red lines)
          and $\alpha=87/39$ (blue lines) we only show the third order
          result. In the insets, we show the relevant second and third
          order expansion coefficients as a function of mass ratio.}
	\label{fig:E_Z_mstar_plot}
\end{figure}

\emph{Plots.---}In Fig.~\ref{fig:E_Z_mstar_plot}, we plot the zero
momentum polaron energy, residue, and effective mass, obtained from
Eqs.~\eqref{eq:mu}, \eqref{ResidueFinal}, and \eqref{eqn:mstar_final},
in the range $-0.3<a/\xi<0.3$ where we expect perturbation theory to
be reliable. 
As was discussed above, one should be careful when $a_{\B}$ approaches zero
since the quasiparticle residue vanishes in this limit.
We have chosen $a_\B/\xi=0.1$ and depict the results in
the case of equal masses ($\alpha=1$), as well as for the mass ratios
$\alpha=39/87$ and $\alpha=87/39$ corresponding to the experimentally
relevant case of a $^{39}$K-$^{87}$Rb mixture.

Consider first the energy. We have $E(0)<0$ [$E(0)>0$] for $a<0$
[$a>0$], corresponding to the attractive (repulsive) branch which are
both described within our perturbation theory. The second order term gives
a significant correction to the energy whereas the third order term is
very small. This is explained in the inset, which shows that the third
order expansion coefficient $B(\alpha)$ is much smaller than the
second order coefficient $A(\alpha)$, except for $\alpha\ll 1$, so
that the third order term is suppressed even when $a/\xi\sim 1$. In
fact, $B(\alpha)\rightarrow 0$ for a very heavy impurity with
$\alpha\rightarrow\infty$~\cite{supmat}. We note that the polaron can form a dimer with a boson  for $a>0$. 
This decay  process is, however, slow in the  perturbative regime considered here, since the molecule is deeply 
bound with a binding energy
$-1/2m_ra^2$.

Consider next the quasiparticle residue $Z$. Here, the third order
term gives a significant correction, increasing $Z$ for the attractive
polaron and decreasing it for the repulsive polaron. As we see from
the inset, this is because the third order coefficient $D(\alpha)$ is
larger than the second order coefficient $C(\alpha)$. Finally, we see
that the third order term gives a large contribution to the effective
mass, decreasing (increasing) it for the attractive (repulsive)
polaron. This is consistent with the inset depicting the expansion
coefficients $F(\alpha)$ and $G(\alpha)$.  We have
$F(\alpha)\rightarrow0$ and $G(\alpha)\rightarrow0$ for
$\alpha\rightarrow\infty$~\cite{supmat}, indicating that the effective
mass equals the bare mass for a very heavy impurity as expected.

Varying $\xi/a_\B$ changes the slope of  $m^*$ and $Z$ as a function of $a/\xi$, but the results are 
qualitatively the same as those depicted in
Fig.~\ref{fig:E_Z_mstar_plot}.  Varying the mass ratio $\alpha$ also changes the relative
weight of the second and third order terms as explained above.  In the
Supplemental Material, we provide the values of
$A(\alpha),\ldots, G(\alpha)$ for $\alpha=0$ and
$\alpha\rightarrow\infty$.  Intriguingly, $C(\alpha)$, $D(\alpha)$,
$F(\alpha)$, and $G(\alpha)$ all diverge for $\alpha\rightarrow0$,
indicating a breakdown of perturbation theory. In this limit, the
atoms in the BEC are much heavier than the impurity, and it would be
interesting to examine how this breakdown is related to the problem of
a mobile impurity interacting with static scatterers.

\emph{Conclusion.---}We  developed a systematic perturbation
theory for the quasiparticle properties of an impurity particle in a
BEC. Analytical results for the energy, residue, and
effective mass were derived, and the energy was shown to contain a third order
logarithmic term, whereas the residue and the effective mass are given by
analytic power series in $a$ up to  third order.  When $a_\B=a$, we obtained the same 
form for the energy as that of a weakly interacting Bose gas, which opens up the 
possibility of detecting corrections to mean-field theory of the  Lee-Huang-Yang and even a fourth order logarithmic type for the first time, using rf spectroscopy.  
The effects of a mass difference between the impurity and the atoms in the
BEC were analyzed throughout.  By  deriving rigorous results for the  quasiparticle properties  of the Bose polaron, our theory   provides a useful benchmark for
approximate many-body theories and for experiments.

\begin{acknowledgments}
  We thank M.~M.~Parish, R.~Schmidt, Y.\ Nishida, and J.\ Arlt for useful discussions.
  R.S.C. and G.M.B. would like to acknowledge the support of the Villum
  Foundation via Grant No. VKR023163.
\end{acknowledgments}

\bibliography{bosepolaron_resubmit}

\widetext
\clearpage
\begin{center}
\textbf{\large Supplemental Materials:
			Quasiparticle Properties of a Mobile Impurity in a Bose-Einstein Condensate}
\end{center}
\setcounter{equation}{0}
\setcounter{figure}{0}
\setcounter{table}{0}
\setcounter{page}{1}
\makeatletter
\renewcommand{\theequation}{S\arabic{equation}}
\renewcommand{\thefigure}{S\arabic{figure}}
\renewcommand{\bibnumfmt}[1]{[S#1]}
\renewcommand{\citenumfont}[1]{S#1}

\section{Renormalization of the contact interaction}
We relate the boson-impurity scattering length to the short-range
potential $V(p)$ as follows. The Lippmann-Schwinger equation yields
the vacuum scattering matrix at vanishing energy in the
center-of-mass frame
\begin{align}
\T_v=\frac1{V(0)^{-1}-\Pi_v},
\label{eq:LS}
\end{align}
in terms of the Fourier transform of the potential at zero momentum
and the vacuum pair propagator $\Pi_v\equiv\int \dif\check k\frac{1}{k^2/2m_r}$. Here we have defined
$\dif\check k\equiv \dif^3k/(2\pi)^3$. At the same time, the vacuum
scattering matrix is related to the boson-impurity scattering length
via
\begin{align}
\T_v=\frac{2\pi a}{m_r}.
\end{align}

Similarly for the boson-boson interaction we have
\begin{align}
\frac{4\pi \ab}{m_{\rm B}}=\frac1{V_{\rm B}(0)^{-1}-\Pi_{\rm B}},
\end{align}
with $\Pi_{\rm B}\equiv\int \dif\check k\frac{1}{k^{2}/m_{\rm B}}$.

\section{Perturbation theory in the scattering length and the  pair propagator}
As explained in the manuscript, to obtain a perturbation theory in the scattering length $a$, we first write down the diagrams in increasing order of the bare interaction $V(q)$. 
We then formally replace $V(q)$ with $ \T_v$ everywhere, which in addition to the obvious substitution $V(q)\rightarrow \T_v$ in the 
expressions for the diagrams, has one more effect:
In diagrams where the ``bare''  forward pair propagator appears when expanding in $V(q)$, we use the regularised pair propagator $\Pi_{11}$ instead when expanding in  $ \T_v$. That is, when 
making the replacement $V(q)\rightarrow \T_v$, we also   make the replacement 
\begin{gather}
-\frac{1}{\beta}\sum_{\omega_\nu}\int \! \dif\check k \, G_{11}(-\k,-i\omega_\nu)G(\k+\p,i\omega_\nu+z)=\int \! \dif\check k\!\left[
  \frac{ u_\k^2(1+f_{\k})}{z - E_\k-\epsilon_{\k+\p}}+\frac{ v_\k^2f_{\k}}
    {z + E_\k-\epsilon_{\k+\p}}\right]\rightarrow\nonumber\\
      \Pi_{11}(p) =\int \! \dif\check k\!
  \left[\frac{ u_\k^2(1+f_{\k})}{z - E_\k-\epsilon_{\k+\p}}+\frac{ v_\k^2f_{\k}}
    {z + E_\k-\epsilon_{\k+\p}} + \frac{2 m_r}{k^{2}}\right].
\label{PairPropagator}
\end{gather}
Here $G(\k,z)=1/(z-\epsilon_k)$ is the non-interacting impurity
Green's function, $i\omega_\nu=i(2\nu+1)T$ is a Fermi Matsubara
frequency, $p=(\p,z)$, $f_\k=[\exp(E_\k/T)-1]^{-1}$ is the boson
distribution function. The term
$2m_r/k^2$ in Eq.~\eqref{PairPropagator} subtracts the contribution
already contained within the vacuum scattering, and acts to
regularize the pair propagator. It comes from inserting the
Lippmann-Schwinger equation \eqref{eq:LS}
 in the ladder approximation for $\T(p)$~\cite{Fetter}. 

For $T=0$ and $p=(0,0)$, Eq.~\eqref{PairPropagator} can be calculated analytically yielding 
\begin{align}
  	\Pi_{11}(0) =
  				\frac{m_\B}{\sqrt{2} \pi^{2} \xi}
  				\frac{\alpha}{1+\alpha}
  				\left[ 1 - \frac{1-\alpha}{1+\alpha} f(\alpha) \right].
				\label{PairPropagatorAna}
				\end{align}				
This gives $\Pi_{11}(0)~=~m_{\B}/2\sqrt{2}\pi^{2}\xi$ for $\alpha=1$.

\subsection{Anomalous and particle-hole propagators}
The anomalous and particle-hole propagators are  
\begin{align}
  \Pi_{12}(p) &=-\frac{1}{\beta}\sum_{\omega_\nu}G_{12}(-\k,-i\omega_\nu)G(\k',i\omega_\nu+z)=\!\int\! \dif\check k
  \left[\frac{u_\k v_\k(1+f_{\k})}{E_\k+\epsilon_{\k'}-z}
    +\frac{u_\k v_\k f_{\k}}{\epsilon_{\k'} - E_\k- z}\right],
\label{AnomalousPropagator}
 \\ 
  \Pi_{22}(p) &=-\frac{1}{\beta}\sum_{\omega_\nu}G_{22}(-\k,-i\omega_\nu)G(\k',i\omega_\nu+z)=\int\! \dif\check k
  \left[\frac{ u_\k^2f_{\k}}{z + E_\k-\epsilon_{\k'}}+
    \frac{v_\k^2(1+f_{\k})}{z - E_\k-\epsilon_{\k'} }\right],
\label{ParticleholePropagator}
\end{align}
where $\k'=\k+\p$,
and $\Pi_{21}(p) =\Pi_{12}(p)$. For 
 $T = 0$ and $p=(0,0)$ we obtain
\begin{align}
  		\Pi_{12}(0)  &=
				\frac{m_\B}{\sqrt{2} \pi^{2} \xi}
				\frac{\alpha}{1 + \alpha}
				f(\alpha), \nonumber
\\
	\Pi_{22}(0) &=
				\frac{m_\B}{\sqrt{2} \pi^{2} \xi}
				\frac{\alpha}{1-\alpha}
				\left[ 1 - f(\alpha) \right].
\label{PairPropagatorsSup}
\end{align}
These expressions  are  well defined in the equal mass limit $\alpha \rightarrow 1$,
where we have
$\Pi_{12}(0)=m_{\B}/2\sqrt{2}\pi^{2}\xi$ and 
$\Pi_{22}(0)=-m_{\B}/6\sqrt{2}\pi^{2}\xi$.

With Eqs.~\eqref{PairPropagatorAna} and \eqref{PairPropagatorsSup}, we have 
analytical expressions for all pair propagators for $p=0$, which is what we need to obtain analytical results for the polaron self-energy 
at zero momentum. 

\section{Logarithmic divergence of \texorpdfstring{$\bm{\Sigma_{3c}}$}{Sigma3c} and \texorpdfstring{$\bm{B(\alpha})$}{C(alpha)}}
Analysing the divergence of $\Sigma_{3c}$, we investigate the integrand in \cref{Crossed}
in the main manuscript for $k \rightarrow \infty$.
We have $u_{\k} \rightarrow 1$,
$v_\k \rightarrow{\T}_{\B}n_0m_{\B}/k^2$,
$E_\k\rightarrow k^2/2m_{\B}$, and
$\Pi_{11}(\k,-E_k)\rightarrow m_r^{3/2}\sqrt{k^2/2m_{\B}+k^2/2M}/\sqrt2\pi$
for $k\rightarrow\infty$, where $M=m+m_\B$. This gives for $p=(\0,0)$
\begin{align}
  \lim_{k\rightarrow\infty}
  \frac{u_{\k} v_{\k}\Pi_{11}(\k,-E_k) }{ E_{\k} 
+ \epsilon_{\bm{\k}} } = \frac{{\T}_{\B}n_0m_{\B}
m_r^2\sqrt{\alpha^2+2\alpha}}{\pi(1+\alpha)k^3},
\label{Limit1}
\end{align}
and 
\begin{align}
  \lim_{k\to\infty}
  \frac{u_{k}^{2} \Pi_{12}(\k,-E_\k) }{ E_{\k} +
  \epsilon_{\bm{\k}} } = 
\frac{{\T}_{\B}n_0Mm_r^2
  (\pi-2\arctan\sqrt{\alpha^2+2\alpha})}{2\pi k^3},
  \label{Limit2}
\end{align}
where we have used 
\begin{align}
I(\alpha)=\int_{\mathrlap{0}}^{\mathrlap{\infty}} \dif x\frac1x\ln\left[\frac{1+x^2+(1+x)^2/\alpha}{1+x^2+(1-x)^2/\alpha}\right]=\pi^2-2\pi\arctan\sqrt{\alpha^2+2\alpha}.
\label{Integral}
\end{align}
Using these limits in \cref{Crossed}, we see that the integrand goes as  $1/k$ for $k\rightarrow\infty$.
Setting the lower and upper limits of the remaining $k$-integral to $1/\xi$ and $1/a^*$ respectively, we obtain \cref{SelfThird1,SelfThird2}.

\section{coefficients for the energy:
			\texorpdfstring{$\bm{A(\alpha})$}{A(alpha)} and \texorpdfstring{$\bm{B(\alpha})$}{B(alpha)}}
We find 
\begin{align}
  A(\alpha) = \frac{2 \sqrt{2}}{\pi} \frac{1}{1 - \alpha} 
  \left[ 1 - \frac{2\alpha^{2}}{1+\alpha} f(\alpha) \right],
  \label{SelfSecond2}
\end{align}
with
$f(\alpha)\equiv
\sqrt{(\alpha+1)/(\alpha-1)}\arctan\sqrt{(\alpha-1)/(\alpha+1)}$
with the definition $\sqrt{-1}=i$.  Note that this expression is well
defined in the limit $\alpha\to1$ corresponding to equal masses, where
we have $A(1)=8\sqrt{2}/3\pi$.

For the logarithmic term, we find 
\begin{align}
B(\alpha)=(1+\alpha)\left[1-\frac{2}{\pi}\arctan
\sqrt{\alpha^2+2\alpha}\right]-\frac{2\sqrt{\alpha^2+2\alpha}}{\pi(1+\alpha)}. 
\label{SelfThird2}
\end{align}
In the case of equal masses, we have $B(1)=2/3-\sqrt{3}/\pi$. 

\section{coefficients for the residue:
			\texorpdfstring{$\bm{C(\alpha})$}{C(alpha)} and \texorpdfstring{$\bm{D(\alpha})$}{D(alpha)}}

The contributions to the residue from the second order diagrams
$\partial_{\omega}\Sigma_{2}(\bm{0},\omega)\vert_{0}$
and the third order ``ladder'' diagrams
$\partial_{\omega}\Sigma_{3b}(\bm{0},\omega)\vert_{0}$
can be calculated analytically, while the  contribution from
the ``crossed'' diagrams has to be evaluated numerically.
We obtain
\begin{align}
	\partial_{\omega}\Sigma_{2}(\bm{0},\omega)\vert_{0}
		&= - \frac{1}{\sqrt{2}\pi} \frac{a^{2}}{a_{\B} \xi}
			\frac{\alpha+1}{\alpha-1}
			\left[ 1 - \frac{2}{\alpha+1} f(\alpha) \right],
\\
	\partial_{\omega}\Sigma_{3b}(\bm{0},\omega)\vert_{0}
		&= \frac{1}{\pi^{2}} \frac{a^{3}}{a_{\B} \xi^{2}}
			\frac{\alpha+1}{\alpha}
			\left\{
				\left(\frac{\alpha+1}{\alpha-1}\right)^{2}
				\left[1 - \frac{2\alpha}{\alpha+1} f(\alpha) \right]^{2}
			-
				\left[1 + \frac{2\alpha}{\alpha+1} f(\alpha) \right]^{2}
			\right\},
\\
	\partial_{\omega}\Sigma_{3c}(\bm{0},\omega)\vert_{0}
		&= \frac{1}{2\pi^{2}} \frac{a^{3}}{a_{\B} \xi^{2}}
			\left( \frac{\alpha+1}{\alpha} \right)^{3}
			I_{\omega}(\alpha)
\end{align}
with
\begin{align}
	I_{\omega}(\alpha) &= \quad \mathclap{\int_{0}^{\infty}} \dif k \left\{
					\frac{k^{2} E_{k}^{(-)}}{ [ E_{k} + k^{2}/\alpha ]^{2} }
					\hspace{.1cm}
								\int_{\mathrlap{0}}^{\mathrlap{\infty}} \dif q
								\int_{\mathrlap{-1}}^{\mathrlap{1}}  \dif t
								\left[
								\frac{q^{2} E_{q}^{(+)}}
									{ E_{k} + E_{q} + (k^{2} + q^{2} - 2 k q t )/\alpha }
												- \frac{2 \alpha}{1 + \alpha}
												\right]
	\right. \nn\\ & \hspace{1.25cm} +
			\frac{k^{2} E_{k}^{(-)}}{ E_{k} + k^{2}/\alpha }
					\hspace{.3cm}
								\int_{\mathrlap{0}}^{\mathrlap{\infty}} \dif q
								\int_{\mathrlap{-1}}^{\mathrlap{1}}  \dif t \;
								\frac{q^{2} E_{q}^{(+)}}
									{ [ E_{k} + E_{q} + ( k^{2} + q^{2} - 2 k q t )/\alpha ]^{2} }
			\nn\\ & \hspace{1.25cm} +
			\frac{k^{2} E_{k}^{(+)}}{ [ E_{k} + k^{2}/\alpha ]^{2} }
								\int_{\mathrlap{0}}^{\mathrlap{\infty}} \dif q
								\int_{\mathrlap{-1}}^{\mathrlap{1}}  \dif t \;
								\frac{q^{2} E_{q}^{(-)}}
									{ E_{k} + E_{q} + ( k^{2} + q^{2} - 2 k q t )/\alpha }
			\nn\\ & \hspace{1.25cm} + \left. 
			\frac{k^{2} E_{k}^{(+)}}{ E_{k} + k^{2}/\alpha }
					\hspace{.3cm}
								\int_{\mathrlap{0}}^{\mathrlap{\infty}} \dif q
								\int_{\mathrlap{-1}}^{\mathrlap{1}}  \dif t \;
								\frac{q^{2} E_{q}^{(-)}}
									{ [ E_{k} + E_{q} + ( k^{2} + q^{2} - 2 k q t )/\alpha ]^{2} }
		\right\}
\end{align}
a dimensionless integral, and 
\begin{align}
	E_{p} = \sqrt{p^{2}(p^{2} + 1)}, \qquad	E_{p}^{(\pm)} = \frac{p^{2} \pm E_{p}}{E_{p}}.
\end{align}
Thus, the coefficients for the residue given by
$Z^{-1} = 1 + C(\alpha) a^{2}/a_{\B}\xi + D(\alpha) a^{3}/a_{\B}\xi^{2}$
are
\begin{align}
	C(\alpha) &= \frac{1}{\sqrt{2}\pi} \frac{\alpha+1}{\alpha-1}
					\left[ 1 - \frac{2}{\alpha} f(\alpha) \right],
\nn \\
	D(\alpha) &=
			-
			\frac{1}{\pi^{2}} \frac{a^{3}}{a_{\B} \xi^{2}}
			\frac{\alpha+1}{\alpha}
			\left\{
				\left(\frac{\alpha+1}{\alpha-1}\right)^{2}
				\left[1 - \frac{2\alpha}{\alpha+1} f(\alpha) \right]^{2}
			-
				\left[1 + \frac{2\alpha}{\alpha+1} f(\alpha) \right]^{2}
			\right\}
			-
			\frac{1}{2\pi^{2}} \left(\frac{\alpha+1}{\alpha}\right)^{3}
			I_{\omega}(\alpha).
\end{align}

\section{coefficients for the effective mass: \texorpdfstring{$\bm{F(\alpha})$}{C(alpha)} and \texorpdfstring{$\bm{G(\alpha})$}{D(alpha)}}
As for the residue, we can calculate the contributions
to the effective mass from the second order diagrams
$\partial_{p^{2}}\Sigma_{2}(\bm{p},0)\vert_{0}$
and the third order ``ladder'' diagrams
$\partial_{p^{2}}\Sigma_{3b}(\bm{p},0)\vert_{0}$
analytically, while the  contribution from
the ``crossed'' diagrams has to be evaluated numerically.
We have
\begin{align}
	2m\partial_{p^{2}}\Sigma_{2}(\bm{p},0)\vert_{0}
		&= \frac{1}{\sqrt{2}\pi} \frac{a^{2}}{a_{\B} \xi}
			\frac{1}{(\alpha-1)^{2}}
			\left[ 1 + \alpha^{2} - \frac{2}{3} \frac{5\alpha^{2}+1}{\alpha+1} f(\alpha) \right],
\\
	2m\partial_{\omega}\Sigma_{3b}(\bm{0},\omega)\vert_{0}
		&= \frac{1}{3\pi^{2}} \frac{a^{3}}{a_{\B} \xi^{2}} \frac{1}{\alpha-1}
			\nn \\ & \hspace{.5cm}
				\times \left[
					\left\{ \frac{3\alpha^{2} - 4\alpha - 1}{\alpha} + \frac{2( 3\alpha^{2} - 2\alpha + 1 )}{\alpha+1} f(\alpha) \right\}
					\left\{ 1 + \frac{2 \alpha}{\alpha+1} f(\alpha) \right\}
			\right. \nn \\ & \left. \hspace{1cm}
				- \left( \frac{\alpha+1}{\alpha-1} \right)^{2}
					\left\{ \frac{3\alpha^{2} + 4\alpha - 1}{\alpha} + \frac{2( 3\alpha^{2} + 2\alpha + 1 )}{\alpha+1} f(\alpha) \right\}
					\left\{ 1 - \frac{2 \alpha}{\alpha+1} f(\alpha) \right\}
				\right],
\\
	2m\partial_{\omega}\Sigma_{3c}(\bm{0},\omega)\vert_{0}
		&= \frac{1}{6\pi^{2}} \frac{a^{3}}{a_{\B} \xi^{2}}
			\left( \frac{\alpha+1}{\alpha} \right)^{3}
			I_{p^{2}}(\alpha)
\end{align}
with
\begin{align}
	I_{p^{2}}(\alpha) &=
		\int_{\mathrlap{0}}^{\mathrlap{\infty}} \dif k \Bigg\{
		\frac{4 k^{3}}{ \alpha [ E_{k} + k^{2}/\alpha ]^{2}}
			\int_{\mathrlap{0}}^{\mathrlap{\infty}} \dif q	\; q^2
			\int_{\mathrlap{-1}}^{\mathrlap{1}} \dif t \;
				\frac{	k - qt 	}
					{\left[E_{k}+E_{q}+(k^2-2 k q t+q^2)/\alpha \right]^2}
		\left( E_{k}^{(-)} E_{q}^{(+)} + E_{k}^{(+)} E_{q}^{(-)} \right)
			\nn\\ & \hspace{1.1cm} -
	\frac{ k^{2} [ 3 E_{k} - k^{2}/\alpha ] }{ [ E_{k} + k^{2}/\alpha ]^{3} }
			\int_{\mathrlap{0}}^{\mathrlap{\infty}} \dif q
			\int_{\mathrlap{-1}}^{\mathrlap{1}} \dif t
				\left[
				\frac{	q^{2} E_{k}^{(-)} E_{q}^{(+)} + q^{2} E_{k}^{(+)} E_{q}^{(-)} 	}
				{ E_{k}+E_{q}+(k^2-2 k q t+q^2)/\alpha }
				-
				\frac{2 \alpha E_{k}^{(-)}}{\alpha+1}
				\right]
			\nn\\ & \hspace{1.1cm} +
	\frac{k^{2} }{ E_{k}+k^2/\alpha }
			\int_{\mathrlap{0}}^{\mathrlap{\infty}} \dif q	\; q^2
			\int_{\mathrlap{-1}}^{\mathrlap{1}} \dif t \;
				\frac{\frac{k^{2}}{\alpha} + \left[ \frac{2 q t}{k} - 3 \right]
						\left[ E_{k} + E_{q} + \frac{q^{2}}{\alpha} \right]
					}
					{ [ E_{k}+E_{q}+(k^2-2 k q t+q^2)/\alpha ]^{3} }
			\left( E_{k}^{(-)} E_{q}^{(+)} + E_{k}^{(+)} E_{q}^{(-)} \right)
		\Bigg\}.
\end{align}
Thus, the coefficients for the effective mass given by
$\frac{m^{*}}{m} = 1 + F(\alpha) a^{2}/a_{\B}\xi + G(\alpha) a^{3}/a_{\B}\xi^{2}$
are
\begin{align}
	F(\alpha) &= -\frac{\sqrt{2}}{3\pi} \frac{3(\alpha+1) - 2(\alpha^{2}+2) f(\alpha)}
						{(\alpha+1)(\alpha-1)^{2}},
\nn \\
	G(\alpha) &=
			\frac{8}{3\pi^{2}}
			\frac{(1+\alpha)^{2}(2+\alpha^{2})
					- 3(\alpha+1)(3\alpha^{2}+1)f(\alpha)
					+ 2\alpha^{2}(\alpha^{2}+5)[f(\alpha)]^{2}}
				{(\alpha-1)(\alpha^{2}-1)^{2}}
	\nn \\ & \hspace{1cm}
		- \frac{1}{2\pi^{2}} \left(\frac{\alpha+1}{\alpha}\right)^{3}
			\left( I_{\omega}(\alpha) + \frac{1}{3} I_{p^{2}}(\alpha) \right).
\end{align}
As stated in the main article the coefficients are well defined in the
limit $\alpha \rightarrow 1$ and the values are stated in \cref{tbl:exp_coeff}.

\section{Expansion coefficients for
\texorpdfstring{$\bm{\alpha=0}$}{alpha=0},
\texorpdfstring{$\bm{\alpha=1}$}{alpha=1} and
\texorpdfstring{$\bm{\alpha=\infty}$}{alpha=infty}}
\begin{table}[h]
	\large
	\centering
	\begin{tabular}{ c c c c c c c }
											&
			\multicolumn{1}{c}{$A(\alpha)$}	&
			\multicolumn{1}{c}{$B(\alpha)$}	&
			\multicolumn{1}{c}{$C(\alpha)$}	&
			\multicolumn{1}{c}{$D(\alpha)$}	&
			\multicolumn{1}{c}{$F(\alpha)$}	&
			\multicolumn{1}{c}{$G(\alpha)$}	\\
		\colrule
			$\alpha \rightarrow 0$
				&	$\frac{2\sqrt{2}}{\pi}$		&	1		&	$\infty$	
				&	$\infty$	&	$\infty$	&	$\infty$
		\\[1ex]
			$\alpha = 1$
			&	$\frac{8\sqrt{2}}{3\pi}$	&	$\frac{2}{3}-\frac{\sqrt{3}}{\pi}$	
				&	$\frac{2\sqrt{2}}{3\pi}$
					&	$\frac{64}{9\pi^{2}} - 0.080$
						&	$\frac{16\sqrt{2}}{45\pi}$
						&	$\frac{448}{135\pi^{2}} - 0.059$
		\\[1ex]
			$\alpha \rightarrow \infty$
				&	$\sqrt{2}$					&	0		&	$\frac{1}{\sqrt{2}\pi}$	
					&	$\frac{2}{\pi} - 0.068$	
					&	0	&	0
		\\[1ex]
	\end{tabular}
	\caption{
		Limiting values of the expansion coefficients for the energy, residue and effective mass. 
		}
	\label{tbl:exp_coeff}
\end{table}

\section{Cancellation of the third order self-energy insertion diagrams}
\begin{figure}[tbp]
\centering
\includegraphics[width=0.2\columnwidth]{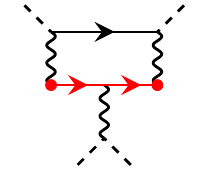}
\caption{Third  order diagram for the self-energy obtained by a self-energy insertion in the 
second order ladder diagram. }
\label{SelfenergiInsertion}
\end{figure}

To illustrate the cancellation of the third order diagrams obtained by self-energy insertions in the second order diagrams, we  give here a specific example. Consider the 
third order diagram depicted in \cref{SelfenergiInsertion}, which is obtained by inserting a self-energy correction in the second order ladder diagram 
shown in Fig.~\ref{2orderdiagrams}(a). For  zero momentum and  temperature, this gives  
\begin{gather}
n_0^2\T_v^3\int \! \dif\check k \,
  \frac{ u_\k^2}{(\omega - E_\k-\epsilon_{\k})^2}.
\label{SelfenergyInsertion}
\end{gather}
When solving $E=\Sigma(E)$ for  the 	impurity energy  to third order in $\T_v$, it is enough to set $\omega=0$ in Eq.\ \eqref{SelfenergyInsertion}, whereas 
we have to insert the first order mean-field energy shift $\omega=n_0\T_v$ in the second order diagrams.  
Inserting this energy shift  in the diagram depicted in Fig.~\ref{2orderdiagrams}(a) and expanding yields
\begin{gather}
n_0\T_v^2\int \! \dif\check k \,
  \frac{ u_\k^2}{n_0\T_v- E_\k-\epsilon_{\k}}
  \simeq n_0\T_v^2\int \! \dif\check k\, \frac{ u_\k^2}{- E_\k-\epsilon_{\k}}\left(1+\frac{ n_0\T_v}{E_\k+\epsilon_{\k}}\right).
  \label{EnergyShiftInsertion}
\end{gather}
This explicitly demonstrates that the third order term in Eq.\ \eqref{EnergyShiftInsertion} cancels the third order diagram given in Eq.\ \eqref{SelfenergyInsertion}.
Note that we have ignored the regularising term in the pair propagator in Eq.\ \eqref{EnergyShiftInsertion}, since it is irrelevant for the present purpose.

\section{Impurity self-energy for an ideal BEC}
For a non-interacting BEC, there is only one impurity Greens function, $G_{11}$, and we furthermore have $u_\k=1$ and $v_\k=0$ for the coherence factors. 
The pair propagator with $p=(\p,\omega)$ becomes the vacuum pair propagator given by~\cite{Fetter} 
\begin{equation}
\Pi_{\text{vac}}(p)=-i\frac{m_r^{3/2}}{\sqrt2\pi}\sqrt{\omega-\p^2/2M},
\end{equation}
 and the anomalous and particle-hole propagators both vanish. It follows that the second order self-energy is 
 \begin{equation}
\Sigma_2(p)=-in_0\T_v^2\frac{m_r^{3/2}}{\sqrt2\pi}\sqrt{\omega-\p^2/2M}.
\end{equation}
The $\sqrt \omega$ dependence means that the residue vanishes for a zero momentum/energy impurity, and there is therefore no well-defined quasiparticle. 

Furthermore,  all third order diagrams vanish for $p=(\0,0)$. The  ladder diagrams depicted in \cref{3orderdiagrams}(a) are zero as they involve the vacuum pair propagator at zero energy/momentum, and/or the anomalous/particle-hole propagators. The crossed diagrams in \cref{3orderdiagrams}(c) are also all zero: Either they contain an integral over the particle-hole or the anomalous propagator which are both zero, or they contain the pair propagator multiplied by a hole propagator, which is zero for a non-interacting BEC. So all 2nd and 3rd order terms vanish for $p=(\0,0)$ for 
 a non-interacting BEC, meaning that mean-field theory  is exact to third order for the energy of a $p=(\0,0)$ impurity.


\end{document}